\documentclass[prd,reprint]{revtex4-1}
\newlength{\picwidth}
 
\usepackage[dvips]{graphicx}
\usepackage{amsmath}

\newcommand{\ern}{\mathcal{E}}
\begin{document}
\renewcommand{\thefigure}{\arabic{figure}}
\title{
 Formal Solution of the Fourth Order Killing equations for  Stationary Axisymmetric Vacuum Spacetimes.}
\author{Jeandrew Brink}
\affiliation{Theoretical Astrophysics, California Institute of Technology, Pasadena, CA 91103 }

\begin{abstract}
An analytic understanding of the geodesic structure around non-Kerr spacetimes will result in a powerful tool that could make the mapping of spacetime around massive quiescent compact objects possible. 
To this end, I present an analytic closed form expression for the components of a the fourth order Killing tensor for Stationary Axisymmetric Vacuum (SAV) Spacetimes. It is as yet unclear what subset of SAV spacetimes admit this solution. The solution is  written in terms of an integral expression involving the metric functions and two specific Greens functions. A second integral expression has to vanish in order for the solution to be exact. In the event that the second integral does not vanish it is likely that the best fourth order approximation to the invariant has been found. This solution can be viewed as a generalized Carter constant providing an explicit expression for the fourth invariant, in addition to the energy, azimuthal angular momentum and rest mass, associated with geodesic motion in SAV spacetimes, be it exact or approximate.   
I further comment on the application of this result for the founding of  a general algorithm for mapping the spacetime around compact objects using gravitational wave observatories.
 
\end{abstract}
\pacs{ }

\maketitle

\section{Introduction}
\label{sec:intro}
During an extreme mass ratio in-spiral (EMRI), a small probe object such as a neutron star spirals into a much more massive compact object commonly held to be a black hole while broadcasting a distinctive gravitational wave (GW) signal. The detection of these GW signals using detectors such as LIGO and LISA could in principle allow us to read off the structure of the massive compact objects \cite{Brownetal,0264-9381-24-17-R01}, much as the head of a gramophone player reads the tune off a long playing record. The feat of so mapping spacetime has been shown to be possible in principle \cite{Ryan}, however an explicit algorithm for doing so in practice for generic objects is still lacking \cite{JdB0}. The GW signal emitted during an inspiral depends on the trajectory of the probe, as well as the propagation of the GW field away from the compact object.   This trajectory can be approximated as a flow through a sequence of geodesics 
and one of the  main stumbling blocks hindering the development of  an algorithm for mapping spacetime is finding an explicit description of the geodesic structure of the spacetime surrounding  the object \cite{JdB0}.    

Our knowledge of the geodesic structure of spacetimes to date encompasses only a subclass of Petrov type D spacetimes found to be separable by Carter \cite{CarterSeparability,JdB2}.  This class of spacetimes, which includes the Kerr metric describing the gravitational field of a rotating black hole, has many special properties \cite{Walker,KressPhd,will:061101,JdB2} which currently undergird our approach to generating EMRI waveforms \cite{WaveFormMachineSteveDrasco}. A key feature of the Carter spacetimes is that they admit a second order Killing tensor $T_{(\alpha_1\alpha_2)}$ which obeys the Killing equations (KE),  $T_{(\alpha_1\alpha_2;\alpha_3)}=0$. As a result, for a particle with momentum $p_\alpha$, the quantity 
$Q_2=T^{(\alpha_1\alpha_2)}p_{\alpha_1} p_{\alpha_2}$
remains constant along its trajectory. This Carter constant $Q_2$, in addition to the constants related to conservation of rest mass  $\mu$,  energy $E$ and azimuthal angular momentum $L_z$ uniquely determine the geodesic structure of the Carter spacetimes.

General SAV spacetimes are of Petrov type I and do not admit second order Killing tensors \cite{JdB3}. However, numerical integrations \cite{Gair, JdB1} performed on SAV spacetimes, exhibited an orbital crossing structure indicative of the existence of a fourth order orbital invariant $Q_4$ quartic in momenta \cite{JdB1}. Such an constant of motion  can be expressed as 
\begin{align}
Q_4=T^{(\alpha_1\alpha_2\alpha_3\alpha_4  )}p_{\alpha_1}p_{\alpha_2}p_{\alpha_3}p_{\alpha_4}, \label{Killv}
\end{align} where the Killing tensor $T$ obeys the fourth order KE 
\begin{align} T_{(\alpha_1\alpha_2 \alpha_3 \alpha_4;\alpha_5)}=0. \label{KEQS}\end{align}
It should be noted that third order Killing tensors need not be considered as  discussed in \cite{JdB1}. 
When written out in full the fourth order KE \eqref{KEQS} represent 56 coupled differential equations, governing  the 35 independent components associated with a totally symmetric tensor $T$. Possibly for this reason, no previous analytic work relating to the solution of fourth order Killing tensors appears to exist. 

This paper provides a non-trivial closed form solution to the fourth order KE \eqref{KEQS} 
in SAV spacetimes. It is the culmination of a serious of papers  \cite{JdB0,JdB1,JdB2,JdB3} investigating the requirements for mapping spacetime and finding a suitable geodesic description for SAV spacetimes that could make such an endeavor tractable. The formalism and variables used in this derivation were originally designed to produce an algebraic check for the existence of  fourth order Killing tensors and are discussed more fully in~\cite{JdB3}. It came as a considerable surprise that one could actually write down the closed form solution presented here.

The analysis is valid for all  SAV spacetimes with two commuting Killing vectors, $\partial_t$ and $\partial_\phi$, represented by means of the Lewis-Papapetrou metric
\begin{align}
ds^2 &=  e^{-2\psi}\left[e^{2\gamma}(d\rho^2+dz^2)+R^2d\phi^2\right]-e^{2\psi}(dt-\omega d\phi)^2. \notag 
\end{align}
The metric functions are entirely determined by a  complex Ernst potential~$\ern$ \cite{Ernst1968} , 
the real part of which is $\Re(\ern)=  e^{2\psi}  $. Line integrals of  $\ern$ determine the functions $\gamma$ and $\omega$. The function $R$ is any harmonic function obeying the equation $R_{zz}+R_{\rho\rho} = 0$ and represents a residual coordinate freedom in the metric.

In Sec. \ref{SEC4OK} a  ``symmetric'' formulation of the fourth order KE derived in \cite{JdB3} is given.  Some of the features which make explicit solution of these equations possible are highlighted.  A constructive derivation of the second order Killing tensor components associated with the Carter spacetimes, given in \cite{JdB2}, indicated that it is possible to write down a formal solution to the equations before determining the explicit SAV metric functions that admit such a solution. A similar feat for the fourth order KE  is performed in Sec. \ref{SECFOURIER}. The resulting solution is given in the form of an integral expression involving the metric functions and appropriately chosen Green's functions.  In addition the requirements for the existence 
 of the solution are also given in integral form.  
 In conclusion the implications of the given solution for a general algorithm for mapping spacetime are discussed in Sec \ref{SECIIII}.

\section{Fourth Order Killing Equations}
\label{SEC4OK}
The KE for SAV spacetimes were carefully analyzed in \cite{JdB3}
and an alternative ``symmetric'' formulation of the fourth order KE derived. This formulation
can be viewed as seeking a solution to four ``interlocking'' fourth order Killing tensor problems for a two-manifold and is entirely equivalent to Eq.~\eqref{KEQS} in the SAV case. The explicit linear transformation between the non-zero fourth order Killing tensor components $T^{(\alpha_1\alpha_2\alpha_3\alpha_4  )}$ and the variables, $ P_{\text{$<$i:j$>$}}$  used in this section  are given in Sec. VIII of \cite{JdB3}. (The transformation  constitutes roughly a page.) The formulation of the KE in terms of $P_{\text{$<$i:j$>$}}$  and the  resulting structure and simplicity  is pivotal in making the problem of solving the large set of KE analytically tractable.

The variable names  $ P_{\text{$<$i:j$>$}}$ are based on the derivative structure imposed by the KE and can be divided into two sets.
 In the first set, the  indices $j=3$ and $j=4$ indicate that the KE fix derivatives with respect to   $\zeta = 1/2(\rho + i z)$ and  $\overline{\zeta} = 1/2(\rho - i z)$ respectively. The index $i$ labels different variables with a fixed derivative structure. 
 The second set with  index  $j=0$ indicates that the KE fully determine the gradient of the variable. 
 There are nine of these variables, $P_{\text{$<$i:0$>$}}$ with \mbox{$i\in\{-4\cdots -1 ,1\cdots 5\}.$} 

The KE governing the derivatives of the 10 variables of the first set are given by the  linear system:
\begin{align}
P_{<i:3>,\zeta }&= -f_{i,\overline{\zeta}} P_{<5:3>}-\frac{1}{4}f_i P_{\text{$<$5:3$>$},\overline{\zeta}}& i\in\{1 \cdots 4\},\notag\\
P_{<i:4>,\overline{\zeta}}&= -f_{i,\zeta } P_{<5:4>}-\frac{1}{4}f_i P_{\text{$<$5:4$>$,$\zeta $}}& i\in\{1 \cdots 4\},\notag\\
P_{\text{$<$5:3$>$,$\zeta $}}&=   P_{\text{$<$5:4$>$,$\overline{\zeta}$}} = 0. 
\label{PKILL3}
\end{align}
 The functions $f_i$  entering these equations are defined in terms of the metric functions as follows:
\begin{align}
f_1&= e^{2 \gamma -2 \psi }=V,&f_2&= \frac{2 e^{2 \gamma }}{3 R^2},\notag\\
f_3&= \frac{2 e^{2 \gamma } \omega }{3 R^2},& f_4&= \frac{e^{2 \gamma } \left(R^2 e^{-4 \psi }-\omega
   ^2\right)}{3 R^2}. \label{METff}
\end{align}
Note that $P_{\text{$<$5:4$>$}}$ in Eq. \eqref{PKILL3} is an analytic function of $\zeta$ and indicates a gauge freedom still present in the metric. Without loss of generality, as was done in the example calculation for second order Killing tensors \cite{JdB2} we can set $P_{\text{$<$5:4$>$}} =P_{\text{$<$5:3$>$}}=1$ 
further simplifying  Eqs. \eqref{PKILL3} to 
\begin{align}
P_{<i:3>,\zeta }&= -f_{i,\overline{\zeta}}& P_{<i:4>,\overline{\zeta}}&= -f_{i,\zeta }& i\in\{1 \cdots 4\}\notag\\
\label{PKILL3b}
\end{align}

To keep the remaining group of KE as succinct as possible and to elucidate the structure found in them more fully, define the operators $fp$ and $\overline{fp}$ to be
\begin{align}
fp(i,j) 
 &= -2( P_{\text{$<$j:3$>$}} f_{\text{i}})_{,\overline{\zeta}}+ f_i  P_{\text{$<$j:3$>$,$\overline{\zeta}$}}\notag\\
\overline{fp(i,j)} &= -2( P_{\text{$<$j:4$>$}} f_{\text{i}})_{,{\zeta}}+ f_i  P_{\text{$<$j:4$>$,$\zeta$}} \label{fp_OPP}
\end{align}
The equations governing the second set of real Killing variables $P_{\text{$<$i:0$>$}}$, whose gradients are fully determined, are given below
\begin{align}
P_{\text{$<$-1:0$>$,$\zeta $}}=&\frac{2}{3} fp(1,1),\notag\\
P_{\text{$<$-2:0$>$,$\zeta $}}=& fp(1,2) + fp(2,1),\notag\\
P_{\text{$<$-3:0$>$,$\zeta $}}=& fp(1,3) + fp(3,1),\notag\\
P_{\text{$<$-4:0$>$,$\zeta $}}=& fp(1,4) + fp(4,1), \notag\\
P_{\text{$<$1:0$>$,$\zeta $}}=&4 fp(2,2),\notag\\
P_{\text{$<$2:0$>$,$\zeta $}}=&- 2(fp(2,3) + fp(3,2)), \notag\\
P_{\text{$<$3:0$>$,$\zeta $}}=& - (fp(2,4) + {fp(4,2)}) + 2fp(3,3), \notag\\ 
P_{\text{$<$4:0$>$,$\zeta $}}=& 4 ( fp(3,4)+{fp(4,3)}),\notag\\
P_{\text{$<$5:0$>$,$\zeta $}}=& 4 fp(4,4), 
\label{KILLPV2a}
\end{align}
The complex conjugates of Eqs. \eqref{KILLPV2a} constitute the remainder of the KE that have to be satisfied and are denoted by  Eqs. CC\eqref{KILLPV2a}

A non-trivial solution of Eqs. \eqref{PKILL3b}, \eqref{KILLPV2a} and CC\eqref{KILLPV2a}, substituted into the transformation from  $P_{<i:j>}$ to $T$ given in Sec. VIII of \cite{JdB3} yields an explicit expression for an invariant of geodesic motion of the form \eqref{Killv}. In Sec. \ref{SECFOURIER} such a solution is proposed. This solution includes the existing class of Carter spacetimes or spacetimes admitting reducible fourth order Killing tensors as well as extending our knowledge to irreducible fourth order Killing tensors. 

\section{Formal Solution of Killing equations via a Fourier Method}
\label{SECFOURIER}
The KE as expressed in Sec.~\ref{SEC4OK} are governed by only  two distinct differential operators. This feature is key to finding an  analytic solution.
 These operators include the Cauchy-Riemann operator $\partial_{\zeta}= \partial_\rho- i \partial_z$ used in Eq.~\eqref{PKILL3b} and on the left-hand side of Eq.~\eqref{KILLPV2a} and the bilinear operator $fp(i,j)$ defined in Eq.~\eqref{fp_OPP} that appears  in various linear combinations in Eq.~\eqref{KILLPV2a}.

Writing down the solution to the KE as given in Sec.\ref{SEC4OK} is a three step process. We first solve for the complex valued functions $P_{<i:3>}$ and $P_{<i:4>}$ that obey Eq. \eqref{PKILL3b}. These results  are substituted into  Eqs. \eqref{KILLPV2a}, which are in turn also solved, yielding an integral expression for $P_{<i:0>}$. The solution satisfies Eqs.  \eqref{PKILL3b} and \eqref{KILLPV2a} and need not be a real valued function. However if it is  actually a solution of the overdetermined set of KE,  Eqs. CC\eqref{KILLPV2a} also have to be satisfied and this provides the constraint that the imaginary part of  $P_{<i:0>}$ 
must vanish. 

There are several analytic solution methods  \cite{Vladimirov1,Vladimirov2,GreensFunctions} that can be used to solve Eq.~\eqref{PKILL3b}, each yielding insight about existence, uniqueness and general behavior of the solution.
The approach followed here is the Fourier method, which has the advantage of making the next step of solving  \eqref{KILLPV2a} easily tractable. In this method the boundary conditions are implicit, the angular variable $z$ is taken to have a range $0\leq z \leq \pi$ with $z= \pi/2$ indicating the equatorial plane and  $\rho$ corresponding to an   asymptotically radial variable extending to infinity. 
We thus take a Fourier series in $z$ and a Fourier integral transform in $\rho$ and define this operation to be the Fourier transform, FT. The FT of a function $h(\tilde{\rho},\tilde{z})$  is denoted by  $ \mathcal{H}(\xi,\eta)$  and defined by 
\begin{align}
  \mathcal{H}(\xi,\eta)= \frac{1}{\pi}\int_{-\infty}^{\infty} \int_0^\pi e^{-i2 \eta \tilde{z}}e^{-2 i \xi \tilde{\rho}}\  h(\tilde{\rho},\tilde{z}) \  d\tilde{\rho} \  d\tilde{z},
\end{align} where $\xi$ is a continuous variable and $\eta$ takes on integer values. The original function is restored by taking the inverse transform, IFT, which can be expressed as
\begin{align}
h(\tilde{\rho},\tilde{z})= \frac{1}{\pi} \sum_{\eta =-\infty}^{\infty}  \int_{-\infty}^{\infty} e^{i2 \eta z}e^{2 i \xi \rho} \ \mathcal{H}(\xi,\eta)
 d\xi .
\end{align}
By applying the FT to Eq. \eqref{PKILL3b}  and recalling that \mbox{$\partial_{\overline{\zeta}} \xrightarrow{FT}  i2\xi-2\eta$} and \mbox{$\partial_{\zeta} \xrightarrow{FT}  i2\xi+2\eta$} we obtain
\begin{align}
\mathcal{P}_{<i:3> }&= - \frac{(2i\xi -2 \eta)}{(2i\xi +2 \eta) } \mathcal{F}_{i},& \mathcal{P}_{<i:4>}&= - \frac{(2i\xi +2 \eta)}{(2i\xi -2 \eta) } \mathcal{F}_{i},\label{FTKILL3}
\end{align}
where $\mathcal{F}_i$ represents the FT of the functions $f_i$ defined in Eq.~\eqref{METff}.
Taking the IFT's of \eqref{FTKILL3} yields an explicit expression for $P_{<i:j>}$. Greater insight into the solution is obtained if we exchange summation and integration and write the solution in terms of a Greens function, $G(\rho,z, \hat{\rho},\hat{z})$. In the case of $P_{\text{$<$i:3$>$}}$  this yields
\begin{align}
P_{\text{$<$i:3$>$}}(\rho,z) =  \frac{1}{\pi}\int_{-\infty}^{\infty} \int_0^\pi f_i(\hat{\rho},\hat{z}) G(\rho,z, \hat{\rho},\hat{z}  ) \  d\hat{\rho} \  d\hat{z},  \label{GP1}
\end{align}
with the Greens function given by
\begin{align}
 G(\rho,z, \hat{\rho},\hat{z}  )= -\frac{1}{\pi} \sum_{\tilde{\eta} =-\infty}^{\infty}  e^{i2 \tilde{\eta} (z- \hat{z}) }\int_{-\infty}^{\infty}e^{2 i \tilde{\xi}( \rho- \hat{\rho})}
 d\tilde{\xi} \notag\\
+\frac{1}{\pi} \sum_{\tilde{\eta} =-\infty}^{\infty}  e^{i2 \tilde{\eta} (z- \hat{z}) }\int_{-\infty}^{\infty}e^{2 i \tilde{\xi}( \rho- \hat{\rho})}\left( \frac{ 4 \tilde{\eta}}{(2i\tilde{\xi} +2 \tilde{\eta}) }\right)
 d\tilde{\xi}. \label{GREENEXP}
\end{align}
Making use of the identities  
\begin{gather}
\int_{-\infty}^{\infty}e^{2 i \tilde{\xi}( \Delta \rho)} d\tilde{\xi}  = \pi \delta( \Delta \rho), \ \
\sum_{\tilde{\eta} =-\infty}^{\infty}  e^{i2 \tilde{\eta} (\Delta z) }= \pi \delta(\Delta z), \notag\\
   \frac{1}{\pi}  \int_{-\infty}^{\infty}  \frac{ e^{2 i \tilde{\xi}\tilde{ \rho}}  }{2( i \tilde{\xi}+n) } d \tilde{\xi} =
  H(\tilde{\rho})e^{-2 n \tilde{\rho}} \ \mbox{if}\ n>0 \ \ \  \mbox{and} \notag\\
\sum_{\tilde{\eta} =1}^{\infty}\tilde{\eta} x^{2\tilde{\eta}}
=\left(x-\frac{1}{x}  \right)^{-2} 
\mbox{if}\ |x|<1,  \label{IDentity}
\end{gather}
where $H(x)$ is the Heaviside step function 
and $\delta(x)$ the Dirac delta function, the expression for the Greens function, given in Eq. \eqref{GREENEXP}, can be reduced to 
\begin{align} 
G(\rho,z, \hat{\rho},\hat{z}  )=  -\pi \delta(z- \hat{z}) \delta(\rho- \hat{\rho}) + \left(\frac{1}{\sinh(  2\hat{\overline{\zeta}}-2\overline{\zeta})}\right)^{2}.\label{GREENF1}
\end{align}
Given expressions \eqref{GP1} and \eqref{GREENF1} for the components $P_{\text{$<$i:3$>$}}(\rho,z)$ we can proceed to the next step of solving Eqs.~\eqref{KILLPV2a} that are governed by the operator $fp(i,j)$ defined in Eq. \eqref{fp_OPP}. All Eqs.~\eqref{KILLPV2a}  can  be written in the form 
\begin{align}
P_{<k,0> ,\zeta}= \sum_n C_n fp(i_n,j_n), \label{EQFTko}
\end{align}
where the integers $n$, $i_n$ and $j_n$ and the rational constants $C_n$ depend on the specific equation being evaluated.
Taking the $FT$ of Eq. \eqref{EQFTko} yields
\begin{widetext}
\begin{align}
\mathcal{P}_{\text{$<$k:0$>$}}(\xi,\eta)=    \sum_n C_n \left( \frac{1}{\pi}\int_{-\infty}^{\infty} \int_0^\pi \  d\tilde{\rho} \  d\tilde{z} \right)  
\left( \frac{1}{\pi}\int_{-\infty}^{\infty} \int_0^\pi d\hat{\rho} \  d\hat{z}\right)
 f_{i_n}(\tilde{\rho},\tilde{z})    e^{-i2 \eta \tilde{z}}e^{-2 i \xi \tilde{\rho}} \notag\\
       \left(   -2  G(\tilde{\rho},\tilde{z}, \hat{\rho},\hat{z}  )  \frac{\left(  2i\xi -2 \eta  \right)}{(2i\xi +2 \eta) } + \frac{\partial}{\partial_{\overline{\tilde{\zeta}}}}\left(  G(\tilde{\rho},\tilde{z}, \hat{\rho},\hat{z}  ) \right) 
  \frac{1}{(2i\xi +2 \eta)} \right)  f_{j_n}(\hat{\rho},\hat{z}).\label{P0111}
\end{align}
We obtain an expression for  $P_{\text{$<$k:0$>$}}$ by taking the $IFT$ of Eq. \eqref{P0111}   and reordering integration and summation to give
\begin{align}P_{\text{$<$k:0$>$}}(\rho,z)
=    \sum_n C_n \left( \frac{1}{\pi}\int_{-\infty}^{\infty} \int_0^\pi \  d\tilde{\rho} \  d\tilde{z} \right)   \left( \frac{1}{\pi}\int_{-\infty}^{\infty} \int_0^\pi \  d\hat{\rho} \  d\hat{z} \right)      f_{i_n}(\tilde{\rho},\tilde{z})   f_{j_n}(\hat{\rho},\hat{z}) K(\rho,z, \tilde{\rho},\tilde{z}, \hat{\rho},\hat{z})
\label{GP2}\end{align}

\end{widetext}

where the ``Greens'' function $K$ was calculated  to be
\begin{align} 
 K(\rho,z, \tilde{\rho},\tilde{z}, \hat{\rho},\hat{z})
&=2G(\tilde{\rho},\tilde{z}, \hat{\rho},\hat{z})G( \rho,z, \tilde{\rho},\tilde{z}) \notag\\
&+ \frac{\partial}{\partial \overline{\tilde{\zeta}}}\left(G(\tilde{\rho},\tilde{z}, \hat{\rho},\hat{z})  \right)  \left(\frac{1}{1-e^{4(\overline{\tilde{\zeta}}-\overline{{\zeta}})}}  \right) \label{GREEN2}
\end{align}
using Eq. \eqref{IDentity}, the expression for $G$, Eq \eqref{GREENF1},  and  the identity 
\begin{align}
\frac{1}{\pi} \sum_{\tilde{\eta} =-\infty}^{\infty}   e^{i2 \tilde{\eta} (\tilde{z}- \hat{z}) }\int_{-\infty}^{\infty} \frac{e^{2 i \tilde{\xi}( \tilde{\rho}- \hat{\rho})}}{(2i\tilde{\xi} +2 \tilde{\eta}) } 
 d\tilde{\xi}&
&= \frac{1}{1- e^{   4\hat{\overline{\zeta}}-4\tilde{\overline{\zeta}}}}
\end{align}
The expressions for the variables $P_{<i,j>}$  given in Eqs.~\eqref{GP1} and~\eqref{GP2} with the Greens functions defined by Eqs.~\eqref{GREENF1} and \eqref{GREEN2} constitute a valid solution to the Eqs. \eqref{PKILL3b} and \eqref{KILLPV2a}. The full set of KE is however over-determined and requires that the additional Eqs.~CC\eqref{KILLPV2a} also be satisfied.  This is synonymous with requiring that the imaginary part of  Eq. \eqref{GP2} is zero. The number of spacetimes that have this property has yet to be determined.

\section{Concluding Remarks}
\label{SECIIII}
This paper provides a formal solution to the fourth order KE for SAV spacetimes. The word formal is used in the sense that a lot of work regarding the evaluations of the integrals presented here still has to be performed before the scope of validity of the solution is determined. To date the expressions given here have been successfully used to predict the Poincare map for the Zipoy-Voorhees (ZV) metric \cite{JdBZV2}. The analytic results agree with well over 500 numerical orbits integrated so far. 

 In the event that the imaginary part of $P_{<i,0>}$  given in \eqref{GP2} is non-zero, (and thus the solution given is not an exact Killing tensor),  the real part of $P_{<i,0>}$ along with $P_{<i,j>}$, $j\in\{3,4\}$ given in Eq. \eqref{GP1} should yield the best fourth order approximation to the invariant  and 
thus provide an analytic handle for describing geodesics in all SAV spacetimes .

In \cite{JdB0} a program for mapping spacetime around compact objects with arbitrary multipole moments was set forth. One of the key stepping stones toward attaining an implementable algorithm was the ability to describe a so-called ``geodesic equivalence class'' between spacetimes that allowed us to explore features of the orbital frequencies and the drift of these frequencies without a-priori knowing what the spacetime was \cite{JdB0}. 
The Green's function formulation of the  solution to the fourth order KE given here allows us to do just that. It provides a rigorous mathematical substructure on which the subsequent algorithmic steps of, translating the GW field away from the compact object, and the detection phase can be built without knowing the specific details of the  spacetime under consideration.

The solution given here  makes explicit 
that missing analytic link, between the geodesic behavior of a particle within a SAV spacetime and the structure of the spacetime itself.
The full exploration of its implications, and further development of the ideas set out in \cite{JdB0} for mapping spacetime will be the subject of future work .

\section{Acknowledgments}
My sincere thanks to Frank Estabrook for many useful discussions. I am also indebted to Tanja Hinderer and Michele Vallisneri for their insightful comments on the manuscript and NITheP, Stellenbosch for their hospitality during part of its preparation. I gratefully acknowledge support from NSF grants PHY-0653653, PHY-0601459, NASA grant NNX07AH06G, the Brinson Foundation and the David and Barbara Groce startup fund at Caltech.

\bibliographystyle{apsrev}

\bibliography{../BholesNemadon}

\begin{thebibliography}{18}
\expandafter\ifx\csname natexlab\endcsname\relax\def\natexlab#1{#1}\fi
\expandafter\ifx\csname bibnamefont\endcsname\relax
  \def\bibnamefont#1{#1}\fi
\expandafter\ifx\csname bibfnamefont\endcsname\relax
  \def\bibfnamefont#1{#1}\fi
\expandafter\ifx\csname citenamefont\endcsname\relax
  \def\citenamefont#1{#1}\fi
\expandafter\ifx\csname url\endcsname\relax
  \def\url#1{\texttt{#1}}\fi
\expandafter\ifx\csname urlprefix\endcsname\relax\def\urlprefix{URL }\fi
\providecommand{\bibinfo}[2]{#2}
\providecommand{\eprint}[2][]{\url{#2}}

\bibitem[{\citenamefont{Brown et~al.}(2007)\citenamefont{Brown, Brink, Fang,
  Gair, Li, Lovelace, Mandel, and Thorne.}}]{Brownetal}
\bibinfo{author}{\bibfnamefont{D.~A.} \bibnamefont{Brown}},
  \bibinfo{author}{\bibfnamefont{J.}~\bibnamefont{Brink}},
  \bibinfo{author}{\bibfnamefont{H.}~\bibnamefont{Fang}},
  \bibinfo{author}{\bibfnamefont{J.~R.} \bibnamefont{Gair}},
  \bibinfo{author}{\bibfnamefont{C.}~\bibnamefont{Li}},
  \bibinfo{author}{\bibfnamefont{G.}~\bibnamefont{Lovelace}},
  \bibinfo{author}{\bibfnamefont{I.}~\bibnamefont{Mandel}}, \bibnamefont{and}
  \bibinfo{author}{\bibfnamefont{K.~S.} \bibnamefont{Thorne.}},
  \bibinfo{journal}{Phys.\ Rev.\ Lett.} \textbf{\bibinfo{volume}{99}},
  \bibinfo{pages}{201102} (\bibinfo{year}{2007}).

\bibitem[{\citenamefont{Amaro-Seoane et~al.}(2007)\citenamefont{Amaro-Seoane,
  Gair, Freitag, Miller, Mandel, Cutler, and Babak}}]{0264-9381-24-17-R01}
\bibinfo{author}{\bibfnamefont{P.}~\bibnamefont{Amaro-Seoane}},
  \bibinfo{author}{\bibfnamefont{J.~R.} \bibnamefont{Gair}},
  \bibinfo{author}{\bibfnamefont{M.}~\bibnamefont{Freitag}},
  \bibinfo{author}{\bibfnamefont{M.~C.} \bibnamefont{Miller}},
  \bibinfo{author}{\bibfnamefont{I.}~\bibnamefont{Mandel}},
  \bibinfo{author}{\bibfnamefont{C.~J.} \bibnamefont{Cutler}},
  \bibnamefont{and} \bibinfo{author}{\bibfnamefont{S.}~\bibnamefont{Babak}},
  \bibinfo{journal}{Class. Quantum Grav.} \textbf{\bibinfo{volume}{24}},
  \bibinfo{pages}{R113} (\bibinfo{year}{2007}).

\bibitem[{\citenamefont{Ryan}(1974)}]{Ryan}
\bibinfo{author}{\bibfnamefont{M.~P.} \bibnamefont{Ryan}},
  \bibinfo{journal}{Phys.\ Rev.\ D} \textbf{\bibinfo{volume}{10}},
  \bibinfo{pages}{1736 } (\bibinfo{year}{1974}).

\bibitem[{\citenamefont{Brink}(2008{\natexlab{a}})}]{JdB0}
\bibinfo{author}{\bibfnamefont{J.}~\bibnamefont{Brink}},
  \bibinfo{journal}{Phys.\ Rev.\ D} \textbf{\bibinfo{volume}{78}},
  \bibinfo{eid}{102001} (\bibinfo{year}{2008}{\natexlab{a}}).

\bibitem[{\citenamefont{Carter}(1968)}]{CarterSeparability}
\bibinfo{author}{\bibfnamefont{B.}~\bibnamefont{Carter}},
  \bibinfo{journal}{Commun.\ Math.\ Phys.} \textbf{\bibinfo{volume}{10}},
  \bibinfo{pages}{280} (\bibinfo{year}{1968}).

\bibitem[{\citenamefont{Brink}(2009{\natexlab{a}})}]{JdB2}
\bibinfo{author}{\bibfnamefont{J.}~\bibnamefont{Brink}},
  \bibinfo{journal}{arXiv:0911.1589 (Phys. Rev. D Submitted)}
  (\bibinfo{year}{2009}{\natexlab{a}}).

\bibitem[{\citenamefont{Walker and Penrose}(1970)}]{Walker}
\bibinfo{author}{\bibfnamefont{M.}~\bibnamefont{Walker}} \bibnamefont{and}
  \bibinfo{author}{\bibfnamefont{R.}~\bibnamefont{Penrose}},
  \bibinfo{journal}{Commun. Math. Phys} \textbf{\bibinfo{volume}{18}},
  \bibinfo{pages}{265} (\bibinfo{year}{1970}).

\bibitem[{\citenamefont{Kress}(1997)}]{KressPhd}
\bibinfo{author}{\bibfnamefont{J.}~\bibnamefont{Kress}}, \bibinfo{journal}{PhD
  Thesis, University of Newcastle}  (\bibinfo{year}{1997}).

\bibitem[{\citenamefont{Will}(2009)}]{will:061101}
\bibinfo{author}{\bibfnamefont{C.~M.} \bibnamefont{Will}},
  \bibinfo{journal}{Phys.\ Rev.\ Lett.} \textbf{\bibinfo{volume}{102}},
  \bibinfo{eid}{061101} (\bibinfo{year}{2009}).

\bibitem[{\citenamefont{Drasco and Hughes}(2006)}]{WaveFormMachineSteveDrasco}
\bibinfo{author}{\bibfnamefont{S.}~\bibnamefont{Drasco}} \bibnamefont{and}
  \bibinfo{author}{\bibfnamefont{S.~A.} \bibnamefont{Hughes}},
  \bibinfo{journal}{Phys.\ Rev.\ D} \textbf{\bibinfo{volume}{D 73}},
  \bibinfo{pages}{024027} (\bibinfo{year}{2006}).

\bibitem[{\citenamefont{Brink}(2009{\natexlab{b}})}]{JdB3}
\bibinfo{author}{\bibfnamefont{J.}~\bibnamefont{Brink}},
  \bibinfo{journal}{arXiv:0911.1595 (Phys. Rev. D Submitted)}
  (\bibinfo{year}{2009}{\natexlab{b}}).

\bibitem[{\citenamefont{Gair et~al.}(2008)\citenamefont{Gair, Li, and
  Mandel}}]{Gair}
\bibinfo{author}{\bibfnamefont{J.~R.} \bibnamefont{Gair}},
  \bibinfo{author}{\bibfnamefont{C.}~\bibnamefont{Li}}, \bibnamefont{and}
  \bibinfo{author}{\bibfnamefont{I.}~\bibnamefont{Mandel}},
  \bibinfo{journal}{Phys.\ Rev.\ D} \textbf{\bibinfo{volume}{77}},
  \bibinfo{pages}{024035} (\bibinfo{year}{2008}).

\bibitem[{\citenamefont{Brink}(2008{\natexlab{b}})}]{JdB1}
\bibinfo{author}{\bibfnamefont{J.}~\bibnamefont{Brink}},
  \bibinfo{journal}{Phys.\ Rev.\ D} \textbf{\bibinfo{volume}{78}},
  \bibinfo{eid}{102002} (\bibinfo{year}{2008}{\natexlab{b}}).

\bibitem[{\citenamefont{Ernst}(1968)}]{Ernst1968}
\bibinfo{author}{\bibfnamefont{F.~J.} \bibnamefont{Ernst}},
  \bibinfo{journal}{Physical Review} \textbf{\bibinfo{volume}{167}},
  \bibinfo{pages}{1175} (\bibinfo{year}{1968}).

\bibitem[{\citenamefont{Vladimirov}(1986)}]{Vladimirov1}
\bibinfo{author}{\bibfnamefont{V.}~\bibnamefont{Vladimirov}},
  \emph{\bibinfo{title}{A Collection of Problems on the Equations of
  Mathematical Physics}} (\bibinfo{publisher}{Mir Publishers},
  \bibinfo{year}{1986}), ISBN \bibinfo{isbn}{3-540-16647-5}.

\bibitem[{\citenamefont{Vladimirov}(1971)}]{Vladimirov2}
\bibinfo{author}{\bibfnamefont{V.}~\bibnamefont{Vladimirov}},
  \emph{\bibinfo{title}{Equations of Mathematical Physics}}
  (\bibinfo{publisher}{Marcel Dekker Inc. 95 Madison Ave. New York 10016},
  \bibinfo{year}{1971}), ISBN \bibinfo{isbn}{0-8247-1713-9}.

\bibitem[{\citenamefont{Butkovskiy}(1982)}]{GreensFunctions}
\bibinfo{author}{\bibfnamefont{A.~G.} \bibnamefont{Butkovskiy}},
  \emph{\bibinfo{title}{Green's Functions and Transfer Functions Handbook}}
  (\bibinfo{publisher}{Ellis Horwood Ltd. Market Cross House, Cooper Street,
  Chichester, West Sussex, PO19 1EB, England}, \bibinfo{year}{1982}).

\bibitem[{\citenamefont{Brink}(2009{\natexlab{c}})}]{JdBZV2}
\bibinfo{author}{\bibfnamefont{J.}~\bibnamefont{Brink}},
  \bibinfo{journal}{Poincare Maps of Static Spacetimes with Equatorial
  Symmetry-Example Zipoy Voorhees Metric.}
  (\bibinfo{year}{2009}{\natexlab{c}}).

\end{thebibliography}

\end{document}